\documentstyle [12pt] {article}

\newcommand{\be}{\begin{equation}}
\newcommand{\ee}{\end{equation}}
\newcommand{\beqs}{\begin{eqnarray}}
\newcommand{\eeqs}{\end{eqnarray}}
\def\nn{\nonumber}
\def\({\left(}
\def\){\right)}
\def\zlZ{{\zeta_2 \over \zeta_1}}
\def\zZl{{\zeta_1 \over \zeta_2}}
\def\zl{\zeta_1}
\def\zZ{\zeta_2}
\def\z{\zeta}
\def\d{\delta}

\def\tD{\tilde{\Delta}}
\def\btD{\tilde{\bar{\Delta}}}
\def\a{\alpha}
\def\b{\beta}

\def\na{\nabla}
\def\tna{\tilde{\nabla}}
\def\btna{\tilde{\bar{\nabla}}}
\def\da{\dot{\alpha}}
\def\db{\dot{\beta}}
\def\pa{\partial}

\def\U{\Upsilon}
\def\tU{\tilde{\Upsilon}}
\def\bU{\bar{\Upsilon}}
\def\btU{\bar{\tilde{\Upsilon}}}

\def\S{\Sigma}
\def\th{\theta}
\def\ni{\noindent}

\begin{document}

\begin{titlepage}

\begin{flushright}
\begin{tabular}{l}
ITP-SB-97-68    \\
PUPT-1746      \\
hep-th/9711135  \\ 
November, 1997 
\end{tabular}
\end{flushright}

\vspace{8mm}
\begin{center}
{\Large\bf Feynman rules in $N=2$ projective superspace II :}

\medskip

{\large\bf Massive hypermultiplets}

\vspace{4mm}
\vspace{16mm}

F.Gonzalez-Rey \footnote{email: glezrey@insti.physics.sunysb.edu},

\vspace{4mm}
Institute for Theoretical Physics  \\
State University of New York       \\
Stony Brook, N. Y. 11794-3840  \\

\bigskip

and \\

\bigskip

R. von Unge \footnote{email: unge@feynman.princeton.edu}\\
{\it Physics Department, Princeton University\\}
{\it Princeton, NJ 08544, USA.\\}

\vspace{20mm}

{\bf Abstract}
\end{center}

Manifest $N=2$ supersymmetric hypermultiplet mass terms can be introduced
in the projective $N=2$ superspace formalism. In the case of complex
hypermultiplets, where the gauge covariantized spinor derivatives have
an explicit representation in terms of gauge prepotentials, it is
possible to interpret such masses as vacuum expectation values of an
Abelian vector multiplet. The duality transformation that relates the
$N=2$ off-shell projective description of the hypermultiplet to the
on-shell description involving two $N=1$ chiral superfields allows us
to obtain the massive propagators of the $N=1$ complex linear fields
in the projective hypermultiplet. The $N=1$ massive propagators of the
component superfields in the projective hypermultiplet suggest a  
possible ansatz for the $N=2$ massive propagator, which agrees
with an explicit calculation in $N=2$ superspace.

\vspace{35mm}

 \end{titlepage}
\newpage
\setcounter{page}{1}
\pagestyle{plain}
\pagenumbering{arabic}
\renewcommand{\thefootnote}{\arabic{footnote}}
\setcounter{footnote}{0}

\section{Introduction}

 A manifestly $N=2$ supersymmetric path integral quantization of $N=2$ 
massless hypermultiplets living in projective superspace
\cite{martin_ulf} has recently been proposed \cite{hyper_feynman}
(for an alternative formulation of $N=2$ superspace, see \cite{harmonic}).
This off-shell representation of $N=2$ supersymmetry contains $N=1$
chiral, complex linear and auxiliary superfields. It can be related to
the traditional on-shell hypermultiplet involving two massless $N=1$
chiral fields via a duality transformation that acts on the complex
linear field of the projective hypermultiplet.

 In this paper we extend the analysis of \cite{hyper_feynman} to
massive hypermultiplets (for the Harmonic superspace version see 
\cite{harm_massive}). As in \cite{hyper_feynman} we first guess the 
massive propagator of the projective hypermultiplet inspired by the 
explicit form of the
massive propagators of its $N=1$ component superfields. The most
efficient way to find these massive $N=1$ propagators is to perform the
duality transformation in the action where we have included couplings
to sources.

 To write the propagators in projective superspace we have to introduce
a central charge in the supersymmetry algebra. It is useful to
identify such a central charge with the expectation value of a
background Abelian vector multiplet. We show that using the centrally
extended spinor derivatives, we can find an ansatz for the $N=2$
propagator. The ansatz takes a simple form which is the naive
generalization of the massless one in \cite{hyper_feynman}. When reduced to
$N=1$ components it gives the correct $N=1$ propagators of the massive
chiral-antichiral, linear-antilinear and auxiliary fields.  Finally
we derive the same $N=2$ propagator directly in $N=2$ superspace by
manipulating the path integral. We end with a list of the Feynman
rules used to calculate Feynman diagrams containing massive
hypermultiplets in projective superspace.

 There is one exception to this result; for {\em real} $O(2p)$
multiplets we cannot express the propagator in projective
superspace. This is related to the fact that we cannot consistently
assign a U(1) charge to a real field.  We may, however, find a
massive $N=2$ propagator if we use a {\em complex} $O(2p)$ multiplet. It
describes two physical hypermultiplets, but otherwise it gives
consistent Feynman rules. In particular, the limit
$p \rightarrow \infty$ of the complex $O(2p)$ multiplet gives the
(ant)arctic multiplet propagator.

\section{Projective Superspace with central charges}
 
We briefly review the basic ideas of $N=2$ projective superspace and
consider the peculiarities introduced by central charges. For a more 
complete review of ordinary $N=2$ projective superspace we refer the 
reader to \cite{martin_ulf},\cite{hyper_feynman}. 

The algebra of $N=2$ supercovariant derivatives with central charges 
in four dimensions is\footnote{ We will use the notation and 
normalization conventions of 
\cite{book}; in particular we denote $D^2 = {1 \over 2} D^\a D_\a$
and $\Box = {1 \over 2} \pa^{\a\da} \pa_{\a\da}$.}

\be
 \{ {\cal D}_{a \a} , {\cal D}_{b \b} \} = q C_{ab} C_{\a \b} \bar{m} \ , \ \  
 \{ \bar{\cal D}^a_{\da} , \bar{\cal D}^b_{\db} \} = - q C^{ab} C_{\da \db} m \ , \ \ 
 \{ {\cal D}_{a \a} , \bar{\cal D}^b_{\db} \} = i \d^b_a \pa_{\a\db } \  .
\label{n2_algebra}
\ee

\ni
where we have included a possible global $U(1)$ charge which will be
useful later. The projective subspace of $N=2$ superspace is parameterized 
by a complex coordinate $\zeta$, and it is spanned by 
the following projective supercovariant derivatives 
\cite{martin_ulf}

\beqs
 \tna_\a (\z) & = & {\cal D}_{1 \a} + \zeta {\cal D}_{2 \a} \nn  \\
 \btna_{\da} (\z) & = & \bar{\cal D}^2_{\da} - \z \bar{\cal D}^1_{\da} \ .
\label{proj_der}
\eeqs

 The conjugate of any object is constructed in this subspace by composing
the antipodal map on the Riemann sphere with hermitian conjugation. To
obtain the 
barred supercovariant derivate we conjugate the unbarred derivative and
we multiply by an additional factor $(-\zeta)$   

\be
- \zeta \overline{\tna_\a (\z)} = \btna_{\da} (\z) \ .
\ee

\ni
The orthogonal combinations 

\be
 \tD_\a = - {\cal D}_{2 \a} + { 1 \over \z } {\cal D}_{1 \a} \; , \;\;
 \btD_{\da} = \bar{\cal D}^1_{\da} + { 1 \over \z } \bar{\cal D}^2_{\da} \ ,
\ee

\ni
and the projective supercovariant derivatives give the following algebra 

\beqs
& \{ \tna , \tna \} = \{ \tna , \btna \} = \{ \tD , \tD \} = 
 \{ \tD , \btD \} = 0 & \label{na_D_algebra1}\\
& \{ \tna_\a , \tD_\b \} = - 2 q C_{\a \b} \bar{m} & \label{na_D_algebra2}\\
& \{ \tna_\a , \btD_{\da} \} = - \{ \btna_{\da} , \tD_\a \} = 
  2i \partial_{\a \da} & \ .
\label{na_D_algebra3}
\eeqs

 For notational simplicity we denote from now on 
${\cal D}_{1 \a} = {\cal D}_\a , {\cal D}_{2 \a} = {\cal Q}_\a$. Superfields
living in $N=2$ projective superspace are annihilated by
the projective supercovariant derivatives (\ref{proj_der}). This
constraints can be rewritten as follows

\be
 \tna_\a \tU = 0 = \btna_{\da} \tU \Longrightarrow
 {\cal D}_\a \tU = - \z \; {\cal Q}_\a \tU  \; \; , \; \; 
 \bar{\cal Q}_{\da} \tU = \z \; \bar{\cal D}_{\da} \tU \ .
\label{proj_const}
\ee

\ni
Manifestly $N=2$ supersymmetric actions have the form 

\be 
 {1 \over 2 \pi i} \oint_C {d \z \over \z} \; d x \; 
 {\cal D}^2 \bar{\cal D}^2 f(\tU, \btU, \z) \ ,
\label{eq-action}
\ee

\noindent
where $C$ is a contour around some point of the complex plane 
that generically depends on the function $f(\tU, \btU,\z)$.

 The superfields obeying (\ref{proj_const}) may be classified 
\cite{martin_ulf} as: i) $O(k)$ multiplets, ii) rational multiplets 
iii) analytic
multiplets. The $O(k)$ multiplet can be expressed as a polynomial in 
$\zeta$ with powers ranging from $0$ to $k$. If we multiply the $O(k)$
multiplet by a factor
$\z^i$ it becomes a polynomial in $\z$ that we will denote as $O(i,i+k)$. 
Rational multiplets are quotients of $O(k)$ multiplets, and analytic 
multiplets are analytic in the coordinate $\z$ on some region of the
Riemann sphere. 

\noindent
For even $k=2p$ we can impose a reality condition on the $O(k)$
multiplet. We use the the name $\eta$ for the real $O(2p)$ 
superfield. The reality condition obeyed by this field can be written 
as

\be
 \overline{\( \tilde{\eta} \over \z^p \)} = { \tilde{\eta} \over \z^p } \ ,
\label{eta_reality}
\ee

\noindent
or equivalently, in terms of coefficient superfields,

\be 
\tilde{\eta}_{2p-n} = (-)^{p-n} \bar{\tilde{\eta}}_n \ .
\ee

 The {\em arctic} multiplet, is the limit $k \rightarrow \infty$ of 
the complex $O(k)$ multiplet. It is therefore
analytic on $\zeta$ around the north pole of the Riemann sphere

\be
\tU = \sum_{n=0}^\infty \tU_n \z^n \ .
\ee

\noindent 
Its conjugate ({\em antarctic}) superfield

\be
\btU = \sum_{n=0}^\infty \btU_n (- {1 \over \z})^n
\ee

\noindent
is analytic around the south pole of the Riemann sphere.

 Similarly, if we consider the self-conjugate projective superfield 
$\tilde{\eta} / \zeta^p$, the real {\em tropical} multiplet can be 
identified with the limit $p \rightarrow \infty$ of this field

\be
V(\zeta) = \sum_{n= -\infty}^{+\infty} v_n \zeta^n \ ;
\ee

\ni
it contains a piece analytic around the north pole of the 
Riemann sphere (though not projective) and a piece analytic around the
south pole:

\be
 V(\z) = V_-(\z) + V_+(\z) \; ; \;\; V_-(\z) = \sum_{n= -\infty}^{-1} 
 v_n \zeta^n + {1 \over 2} v_0 \; , \;\;  V_+(\z) = {1 \over 2} v_0
 +  \sum_{n=1}^{+\infty} v_n \zeta^n \ . 
\label{splitting}
\ee

\ni
The reality condition in terms of its coefficient superfields is the 
following

\be 
v_{-n}= (-)^n \bar{v_n} \ .
\ee

 The constraints obeyed by multiplets living in projective superspace 
(\ref{proj_const}) can be written in terms of their coefficients

\be
 {\cal D}_\a \tU_{n+1} = - {\cal Q}_\a \tU_n  \;\; , \;\;
 \bar{\cal D}_{\da} \tU_n = \bar{\cal Q}_{\da} \tU_{n+1} \ .
\ee

\noindent
Such constraints imply that the lowest order coefficient superfield 
of any multiplet is antichiral in $N=1$ superspace, and the next to 
lowest order obeys a modified antilinearity constraint. The same 
constraints imply that the highest order coefficient superfield is
chiral in $N=1$ superspace and the next to highest order obeys a 
modified linearity constraint

\beqs
& & {\cal D}_\a \tU_0 = 0 \; , \;\; {\cal D}^2 \tU_1 = q \bar{m} \tU_0 
 \; , \;\; \bar{\cal D}_{\da} \tU_k = 0 \; , \;\; 
 \bar{\cal D}^2 \tU_{k-1} = q m \tU_k \label{n1_constr}  \\
& & \bar{\cal D}_{\da} \btU_0 = 0 \; , \;\; 
 \bar{\cal D}^2 \btU_1 = - q m \btU_0 \; , \;\; {\cal D}_\a \btU_k = 0 
 \; , \;\; {\cal D}^2 \btU_{k-1} = - q \bar{m} \btU_k  \ . \nn
\eeqs

\ni
In the case of complex $O(k)$ hypermultiplets these highest
and lowest order superfields are not conjugate to each other, and
the complex multiplet describes twice as many physical degrees of freedom as
the real one \cite{hyper_feynman}.

 In the case of the real projective multiplet there is no lowest or
highest order coefficient, and therefore none of the
coefficient superfields is constrained in $N=1$ superspace.

\section{Massive hypermultiplet in $N=1$ superspace}

\subsection{N=1 duality of the off-shell and on-shell hypermultiplet 
              descriptions}

 To illustrate how the central charges arise in the algebra of $N=2$
supercovariant derivatives and understand the role they play in the
duality of the on-shell and off-shell realizations, we begin considering
hypermultiplets that live in a projective superspace without central 
charges. The supercovariant derivatives in this space obey the following 
algebra

\be
 \{ D_{a \a} , D_{b \b} \} = 0 \ , \ \  
 \{ D_{a \a} , \bar{D}^b_{\db} \} = i \d^b_a \pa_{\a\db } \  .
\label{noc_n2_algebra}
\ee 

\ni
The kinetic action of a massless (ant)arctic hypermultiplet living in 
this space is

\be 
S_0 = \int {d^4x D^2 \bar{D}^2} \oint {d\zeta \over 2 \pi i \zeta} \; \; 
 \bU \U \ .
\label{eq-Lagrangian2}
\ee

 The component form of this action and the $N=1$ duality
transformation that converts it into the well known action of the 
hypermultiplet realizing $N=2$ supersymmetry on shell,
have been discussed in ref. \cite{hyper_feynman}. If we include
a coupling with an external Abelian vector multiplet living in
$N=2$ projective superspace \cite{martin_ulf2} we have

\be 
S = \int d^4x \, d^4 \th \oint {d\zeta \over 2 \pi i \zeta} \; \; 
 \bU e^{V} \U \ ,
\label{eq-Lagrangian3}
\ee

\ni
where $V(\z)$ is a real tropical multiplet. This action is invariant
under $N=2$ supersymmetry transformations and also under $U(1)$ gauge
transformations 

\be
 \bU' = e^{i q \bar{\Lambda}} \bU \;\;\; , \;\;\; (e^V)' = 
 e^{i \bar{\Lambda}} e^V e^{- i \Lambda}  \;\;\; , \;\;\; 
 \U' = e^{i q \Lambda}\U  \ ,
\label{gauge_sym}
\ee
 
\ni
when we assign $U(1)$ charges $q(\U)=1, q(\bar{\U})=-1$. The most 
general gauge parameter $\Lambda(\z)$ is a projective arctic multiplet. 
Splitting the tropical multiplet into the 
antarctic piece and the arctic piece (\ref{splitting})

\be
 e^{V(\z)} = e^{V_- (\z)} e^{V_+ (\z)} \ ,
\ee

\noindent
we can define gauge covariantized projective spinor derivatives  
\cite{martin_ulf2} :

\be
 \tilde{\na}_\a = \na_\a + q (\na_\a V_-) = \na_\a - q (\na_\a V_+)  
 = {\cal D}_\alpha + \zeta {\cal Q}_\alpha
\ee

\ni
and
 
\be
 \tilde{\bar{\na}}_{\da} = \bar{\na}_{\da} + q (\bar{\na}_{\da} V_-) = 
 \bar{\na}_{\da} - q (\bar{\na}_{\da} V_+)
 = \bar{\cal Q}_{\da} - \zeta \bar{\cal D}_{\da} \ ,
\ee

\ni
which annihilate the {\em covariantly projective} (ant)arctic multiplet

\be
 \tilde{\U} = e^{V_+} \U \; , \;\;\; 
 \tilde{\bar{\U}} = e^{V_-} \bar{\U} \ .
\label{covariant_arctic}
\ee

\ni
This covariantly projective hypermultiplet transforms only under the
residual $U(1)$ gauge symmetry,

\be
 \tU' = e^{V_+'} \U' = e^{V_+}
 e^{{i \over 2} (\bar{\Lambda}_0 + \Lambda_0) - i \Lambda} e^{i \Lambda} \U 
 = e^{{i \over 2} (\bar{\Lambda}_0 + \Lambda_0)} \tU \ ,
\label{residual1}
\ee

\ni
and correspondingly its conjugate transforms with the opposite 
$U(1)$ charge

\be
 \btU' = e^{- {i \over 2} (\bar{\Lambda}_0 + \Lambda_0)} \btU \ .
 \label{residual2}
\ee

\ni
The interacting hypermultiplet action (\ref{eq-Lagrangian3}) can be 
written as the kinetic action of a covariantly projective (ant)arctic
multiplet, very much like we do with a covariantly chiral $N=1$
superfield

\be
 S = \int d^4x \, d^4 \th \oint {d\zeta \over 2 \pi i \zeta} \; \; 
 \bU e^{V} \U = \int d^4x \, d^4 \th 
 \oint {d\zeta \over 2 \pi i \zeta} \; \; \btU \tU \ .
\label{cov_action}
\ee

 The anticommutators of gauge covariantized spinor derivatives  

\be
 \{ {\cal D}_\a , {\cal Q}_\b \} = q C_{\a \b} D^2 v_1 \; , \; \;
 \{ \bar{\cal D}_{\da} , \bar{\cal Q}_{\db} \} = q C_{\da \db} 
 \bar{D}^2 v_{-1}
\label{central_charge}
\ee

\ni
are proportional to the $N=2$ gauge field strength $\bar{W} = i 
D^2 v_1$ and its conjugate . These scalar field strengths and the 
spinorial $W_\a=\bar{D}^2 D_\a v_0$ are the only gauge invariant 
fields contained in the tropical multiplet \cite{martin_ulf2}. 
The constrained $N=1$ superfields in $\tU$ are a covariantly antichiral 
field $\tU_0$ and a superfield obeying\footnote{ An interesting  
generalization of this constraint was proposed long time ago for $N=1$
theories with complex linear multiplets in \cite{deo_gates}. In that 
reference it was pointed out that the mass terms of such fields mix them 
with chiral multiplets.} ${\cal D}^2 \tU_1 = - i q \bar{W} \tU_0$. 
A nonzero {\em v.e.v.} of the scalar field strength introduces a 
central charge in the algebra of $N=2$ gauge covariantized spinor 
derivatives. Therefore, to restrict our analysis to the massive 
hypermultiplet we consider only the {\em v.e.v.} of the 
gauge multiplet 

\beqs 
 V & = & \bar{m} { (\th_2 - \th_1 \z)^2 \over 2 \z } - m 
 { (\bar{\th}^1 + \bar{\th}^2 \z)^2 \over 2 \z }  \\
& = & \( \th_2^2 \bar{m} - (\bar{\th}^1)^2 m \) {1 \over \z} - 
 \( \th_2 \th_1 \bar{m} + \bar{\th}^2 \bar{\th}^1 m \) + 
 \( \th_1^2 \bar{m} - (\bar{\th}^2)^2 m \) \z  \ , \nn  
\eeqs

\ni
and to distinguish fields that obey the constraints (\ref{proj_const})
with central charges in the algebra of spinorial derivatives from
those that obey the same constraints without central charges, we 
refer to the former as {\em covariantly projective} and to the latter
as {\em ordinary projective}.

 Now that we have recovered the algebra (\ref{n2_algebra}) of $N=2$ 
spinorial derivatives with central charges acting on a hypermultiplet
which has $q(\tU)=1, q(\btU)=-1$, we want to perform the duality 
transformation that in the massless
case exchanges the $N=1$ complex linear field by a chiral one
\cite{hyper_feynman}. The action (\ref{cov_action}) in components is

\be
 S = \int d^4x \, d^4 \th \;\; \( \btU_0 \tU_0
 - \btU_1 \tU_1 + \sum_{n=2}^{+\infty} (-)^n \btU_n \tU_n \) \ .
\ee

 The path integral of the theory may be rewritten using a parent action
that includes Lagrange multipliers imposing the modified $N=1$ linearity 
constraint on $\btU_1$ (\ref{n1_constr}) and the corresponding constraint 
on its conjugate

\be
 S = \int d x \; d^4 \th \; \;  \left[ \btU_0 \tU_0 - \btU_1 \tU_1 + 
 \btU_2 \tU_2 + \dots + Y (\bar{\cal D}^2 \btU_1 - m \btU_0) 
 + \bar{Y} ({\cal D}^2 \tU_1 - \bar{m} \tU_0) \right] \ .  
\ee

\noindent
We can integrate out the unconstrained field $\tilde{\U}_1$ to get the
kinetic term of a chiral field $\tilde{\phi} = \bar{\cal D}^2
Y$, a chiral mass term mixing $\btU_0$ and $\tilde{\phi}$
(recall that $\btU_0$ is chiral) and its complex conjugate, plus
auxiliary field terms that decouple   

\be
 S_{dual} = \int d x \, d^4 \th \; \; ( \btU_0 \tU_0 + 
 \bar{\tilde{\phi}} \tilde{\phi} + \dots ) 
 - \int d x \, d^2 \th \; \tilde{\phi} \, m \btU_0 
 - \int d x \, d^2 \bar{\th} \; \; \bar{\tilde{\phi}} \, \bar{m} \tU_0 \ . 
\label{massive_action}
\ee

\ni
This provides the traditional description of massive hypermultiplets 
in terms of two $N=1$ chiral scalars 

 We can also introduce mass terms for a complex $O(2p)$
hypermultiplet\footnote{As noted in \cite{hyper_feynman} the complex
$O(k)$ multiplets include chiral ghosts for odd $k$, and although
their $N=2$ propagator is consistent with the general form derived
below, we will restrict ourselves to the case of physical fields.}. 
Its phase rotates under the gauge transformations 
(\ref{residual1}-\ref{residual2}) with the same $U(1)$ charge 
assignment as the (ant)arctic multiplet. To illustrate this we give 
the following explicit construction of the covariantly projective 
$O(2p)$ multiplet.

As was shown in \cite{martin_ulf2}, it is possible to make a partial
$N=2$ {\em supersymmetric} Lindstr\"{o}m-Ro\v{c}ek gauge choice for
the vector multiplet. The basic idea is to truncate the $\z$ expansion
of $V$ by gauging away all but a finite number of its components while
preserving $N=2$ supersymmetry. In this gauge we set $v_i^{LR}=0 \: 
\forall \: i \neq -1,0,1$ and we gauge away all of the prepotential 
$v_1$ except the pieces $D^{2}v_{1}$ and $\bar{Q}^2 v_{1}$, 
leaving us with a $v_1^{LR}$ which is quadratic in Grassmann 
coordinates \cite{martin_ulf2}. In this gauge it is straightforward to see
that a covariantly projective hypermultiplet constructed from an
ordinary projective complex $O(2p-2)$ field $\rho$ is a complex $O(2p)$
multiplet

\be
 \tilde{\rho} = e^{V_+^{LR}} \rho = e^{v_0^{LR} \over 2} 
 \( 1 + v_1^{LR} \z + (v_1^{LR})^2 \z^2 \) \rho \ .
\label{covariant_2p}
\ee

\ni
A generic gauge transformation (\ref{gauge_sym}) maps the gauge 
multiplet $V^{LR}$ into a real tropical multiplet with an infinite 
number of nonzero
coefficients and the finite multiplet $\rho$ transforms into an arctic
multiplet\footnote{This makes perfect sense if we regard the $O(2p-2)$
multiplet as a special (ant)arctic multiplet in a particular gauge, 
{\em i.e.} we are working with the space of $O(\infty)$ polynomials which
contains the subspace of $O(2p-2)$ polynomials.}. At the same time 
the $O(2p)$ multiplet $\tilde{\rho}$ remains a $O(2p)$ multiplet after 
the gauge transformation, as follows from equation (\ref{residual1}). 

Since this multiplet is a special case of the (ant)arctic multiplet, 
we use the same notation $\tU$ and $U(1)$ charge assignment for both. 
The covariantly projective complex $O(2p)$ multiplet therefore obeys 
the constraints (\ref{n1_constr}) with $q=1$ and its conjugate with 
$q=-1$. The action of the free complex $O(2p)$ multiplet written with 
Lagrange multipliers is
   
\beqs
 S & = & \int d x \; d^4 \th \; \left[ \btU_0 \tU_0 - 
 \btU_1 \tU_1 + \dots - \btU_{2p-1} \tU_{2p-1} + \btU_{2p} \tU_{2p} 
 + Y (\bar{\cal D}^2 \btU_1  - m \btU_0) \right. \nn   \\
& & \left. \qquad + \bar{Y} ({\cal D}^2 \tU_1 - \bar{m} \tU_0) + 
 X (\bar{\cal D}^2 \tU_{2p-1} - m \tU_{2p})  
 + \bar{X} ({\cal D}^2 \btU_{2p-1} - \bar{m} \btU_{2p} ) \right] \ .
\eeqs

\ni
Integrating the unconstrained fields as before we obtain the free
action of two copies of massive chiral hypermultiplets plus auxiliary 
fields

\beqs
S_{dual} & = & \int d x \; d^4 \th \; \; ( \btU_0 \tU_0 + 
 \bar{\cal D}^2 Y {\cal D}^2 \bar{Y} + \dots + 
 \bar{\cal D}^2 X {\cal D}^2 \bar{X} ) \nn \\
& & - \int d x \; d^2 \th \;m ( \bar{\cal D}^2 Y \btU_0 + 
 \bar{\cal D}^2 X \tU_{2p} ) - \int d x \; d^2 \bar{\th}  \; 
 \bar{m} ( {\cal D}^2 \bar{Y} \tU_0 + {\cal D}^2 \bar{X} \btU_{2p}) \ .
\eeqs

\subsection{Massive $N=1$ propagators of chiral, linear and auxiliary
            fields}

 To derive the propagators of the $N=1$ superfields contained in the
massive (ant)arctic hypermultiplet, we introduce the following (ant)arctic
source living in projective superspace 

\beqs
 S_0 + S_j & = & \int {d^4x D^2 \bar{D}^2} \;\; \oint 
 {d \zeta \over 2 \pi i \zeta} \btU \tU + { \bar{j} \over \z^2 } \tU   
 + \z^2 \btU j  \label{n1_source_act}    \\
& = & \int {d^4x D^2 \bar{D}^2} \;\; \( \btU_0 \tU_0 - \btU_1 \tU_1  
  + \btU_2 \tU_2 + \dots + \bar{j}_2 \tU_0 + \btU_0 j_2 \right.  \nn \\
& & \qquad \qquad \left. + \bar{j}_3 \tU_1 + \btU_1 j_3 + 
 \bar{j}_4 \tU_2 + \btU_2 j_4 + \dots \) \ . \nonumber  
\eeqs

\ni
After introducing Lagrange multipliers that multiply the linearity
constraints modified by central charges, we integrate out
the unconstrained field $\tU_1$ as we did before. We
obtain the free action of two massive chiral fields and auxiliary 
fields coupled to $N=1$ unconstrained sources plus a term quadratic 
in sources

\beqs
 S_0 + S_j &=&- \int d x \; d^2 \th \;\; m \tilde{\phi} \btU_0 
 - \int d x \; d^2 \bar{\th} \; \; \bar{m} \bar{\tilde{\phi}} \tU_0    \\ 
& & + \int d x \; d^4 \th \; \; \( \btU_0 \tU_0 + 
 \bar{\tilde{\phi}} \tilde{\phi} + \btU_2 \tU_2 + \dots 
 + \bar{j}_2 \tU_0 + \btU_0 j_2 \right. \nn \\
& & \qquad \qquad \qquad \left. + \bar{\tilde{\phi}} j_3 + 
 \bar{j}_3 \tilde{\phi} + \bar{j}_4 \tU_2 + \btU_2 j_4 + \dots 
 + \bar{j}_3  j_3 \) \nn \ .
\eeqs
 
 The path integral of this free theory can be obtained using standard
$N=1$ superspace technology to integrate out the chiral fields: we
rewrite the chiral mass terms as nonchiral integrals and we insert chiral
and antichiral projectors in the corresponding sources 

\beqs
S & = & \int dx \; d^4 \th \;
    {1 \over 2} \( \btU_0, \tU_0, \tilde{\phi}, \bar{\tilde{\phi}} \)
 \( \begin{array}{cccc}
       0   & 1      & {-m {\cal D}^2 \over \Box} &       0             \\
       1   &    0   &   0   &  { -\bar{m} \bar{\cal D}^2 \over \Box} \\
       {-m {\cal D}^2 \over \Box} & 0 & 0 &       1      \\
       0 & { -\bar{m} \bar{\cal D}^2 \over \Box} & 1 & 0 
    \end{array} \)
 \( \begin{array}{c}
       \btU_0 \\  \tU_0 \\  \tilde{\phi} \\ \bar{\tilde{\phi}}  
     \end{array} \) 
 + \btU_2 \tU_2 + \dots  \nn \\ \nn \\
& & \qquad + \( \btU_0, \tU_0, \tilde{\phi}, \bar{\tilde{\phi}} \) 
 \( \begin{array}{c}
     { {\cal D}^2 \bar{\cal D}^2 \over \Box} j_2  \\
    { \bar{\cal D}^2 {\cal D}^2 \over \Box} \bar{j}_2          \\
    { {\cal D}^2 \bar{\cal D}^2 \over \Box} \bar{j}_3  \\
    { \bar{\cal D}^2 {\cal D}^2 \over \Box} j_3
   \end{array} \) +  j_3 \bar{j}_3 + \btU_2 j_4 +
   \bar{j}_4 \tU_2 + \dots 
\eeqs

\ni 
Inverting the mass-kinetic matrix we can complete squares on the chiral 
and antichiral superfields. The resulting path integral is 

\beqs
 ln Z_0[j_i, \bar{j}_i] = - \int dx \; d^4 \th \;
   \sum_{n=3}^{+\infty} (-)^n j_n \bar{j}_n  
 \qquad \qquad \qquad \qquad \qquad \qquad \qquad \qquad \qquad & & 
\eeqs
 
\beqs 
+ { 1 \over 2} \( j_2 { \bar{\cal D}^2 {\cal D}^2 \over \Box}, 
 \bar{j}_2 { {\cal D}^2 \bar{\cal D}^2 \over \Box},
 \bar{j}_3 { \bar{\cal D}^2 {\cal D}^2 \over \Box}, 
 j_3 { {\cal D}^2 \bar{\cal D}^2 \over \Box} \) 
 \( \begin{array}{cccc}
   0 & {- \Box \over \Box - m \bar{m}} & 
   {- \bar{m} \bar{\cal D}^2 \over \Box - m \bar{m} } & 0 \\
   {- \Box \over \Box - m \bar{m}} & 0 & 0 & 
                     {-m {\cal D}^2 \over \Box - m \bar{m} } \\
   {- \bar{m} \bar{\cal D}^2 \over \Box - m \bar{m} } & 0 & 0 & 
        {- \Box \over \Box - m \bar{m}} \\
   0 & {- m {\cal D}^2 \over \Box - m \bar{m} } & 
    {-\Box \over \Box - m \bar{m}} & 0 
 \end{array} \)
 \( \begin{array}{c}
    { {\cal D}^2 \bar{\cal D}^2 \over \Box} j_2  \\
    { \bar{\cal D}^2 {\cal D}^2 \over \Box} \bar{j}_2   \\
    { {\cal D}^2 \bar{\cal D}^2 \over \Box} \bar{j}_3 \\
    { \bar{\cal D}^2 {\cal D}^2 \over \Box} j_3  
   \end{array} \) .        \nn 
 & & \nn
\eeqs

As in the massless case, the only novelty is the term quadratic in sources
which gives an essential contribution to the linear-antilinear
propagator. The only nonvanishing propagators connecting the $N=1$ 
superfields of the massive (ant)arctic hypermultiplet are the following
(we omit their complex conjugates)

\be
 \langle \tilde{\Upsilon}_0 (1) \bar{\tilde{\Upsilon}}_0 (2) \rangle =
 - { {\cal D}^2 \bar{\cal D}^2  \over \Box - m \bar{m} } \d^4 (\th_{12})
 \d^4 (x_{12}),
\label{prop_first}
\ee    

\be
\langle \tilde{\Upsilon}_1 (1) \bar{\tilde{\Upsilon}}_1 (2) \rangle =
 \(1 - { \bar{\cal D}^2 {\cal D}^2 \over \Box - m \bar{m} } \) 
 \d^4 (\th_{12}) \d^4 (x_{12}),
\ee    

\be
\langle \tilde{\Upsilon}_{n>1} (1) \bar{\tilde{\Upsilon}}_{n>1} (2) \rangle =
 (-)^{n+1} \d^4 (\th_{12}) \d^4 (x_{12}),
\ee    

\be
\langle \tilde{\Upsilon}_1 (1) \bar{\tilde{\Upsilon}}_0 (2)\rangle =
 - {\bar{m} \bar{\cal D}^2 \over \Box - m \bar{m} } \d^4 (\th_{12}) 
 \d^4 (x_{12}),
\ee    

\be
\langle \tilde{\Upsilon}_0 (1) \bar{\tilde{\Upsilon}}_1 (2) \rangle =
 - {m {\cal D}^2 \over \Box - m \bar{m} } \d^4 (\th_{12}) \d^4 (x_{12}).
\ee    

In the case of the complex $O(2p)$ hypermultiplet we have in addition
propagators for the second copy of physical $N=1$ superfields.

\be
\langle \tU_{2p-1} (1) \btU_{2p-1} (2) \rangle =
 \(1 - { {\cal D}^2 \bar{\cal D}^2 \over \Box - m \bar{m} } \) 
 \d^4 (\th_{12}) \d^4 (x_{12}),
\ee    

\be
\langle \tU_{2p} (1) \btU_{2p-1} (2) \rangle =
 - {\bar{m} \bar{\cal D}^2 \over \Box - m \bar{m} } \d^4 (\th_{12}) 
 \d^4 (x_{12}),
\ee

\be
\langle \tU_{2p-1} (1) \btU_{2p} (2) \rangle =
 - {m {\cal D}^2 \over \Box - m \bar{m} } \d^4 (\th_{12}) \d^4 (x_{12}).
\ee  

\be
 \langle \tU_{2p} (1) \btU_{2p} (2) \rangle =
 - { \bar{\cal D}^2 {\cal D}^2 \over \Box - m \bar{m} } \d^4 (\th_{12})
 \d^4 (x_{12}),
\label{prop_last}
\ee

\section{Massive hypermultiplet in $N=2$ superspace}

\subsection{Ansatz for the $N=2$ massive propagator}
 
 Now that we have the massive propagators of the $N=1$ component 
fields, we may try to guess the form of the $N=2$ massive propagator 
for the complex $O(2p)$ multiplet. In $N=1$ components it must be of the
following form

\beqs
 \langle \tU (1) \; \btU (2) \rangle \mid_{\th_{\a}^2=0} & = &
 \langle \tU_0 (1) \btU_0 (2) \rangle \ + 
 \zl \ \langle \tU_1 (1) \btU_0 (2) \rangle -
 {1 \over \zZ} \langle \tU_0 (1) \btU_1 (2) \rangle    \\
& & + \(\zZl\)  \langle \tU_1 (1) \btU_1 (2) \rangle + 
 \sum_{n=2}^{2p-2} \(-\zZl\)^n \langle \tU_n (1) \btU_n (2) \rangle
    	\nn  \\ 
& & + \(\zZl\)^{2p-1} \langle \tU_{2p-1} (1) \btU_{2p-1} (2) \rangle + 
 {1 \over \zZ} \(\zZl\)^{2p-1} \langle \tU_{2p-1} (1) \btU_{2p} (2) \rangle
 \nn    \\
& & - \zl \(\zZl\)^{2p-1} \langle \tU_{2p} (1) \btU_{2p-1} (2) \rangle +
 \(\zZl\)^{2p} \langle \tU_{2p} (1) \btU_{2p} (2) \rangle  \nn 
\eeqs

Just as we did in the massless case, we substitute the expressions
(\ref{prop_first} - \ref{prop_last}) and the hypermultiplet propagator is

\beqs
\lefteqn{\langle \tU (1) \; \btU (2) \rangle \mid_{\th_{\a}^2=0} \, =  
 \( \(\zZl\)^{2p-1} - 1 \) \times }\\ \nn \\
& & \( { {\cal D}^2 \bar{\cal D}^2 \over \Box - m \bar{m}}
 + \zl {\bar{m} \bar{\cal D}^2 \over \Box - m \bar{m} } 
 - {1 \over \zZ} {m {\cal D}^2 \over \Box - m \bar{m} } 
 - \(\zZl\) {\bar{\cal D}^2 {\cal D}^2 \over \Box - m \bar{m}} + 
 {\zl \over \zZ-\zl} \) \d^4 (\th_{12}) \d^4 (x_{12}) \ .\nn 
\eeqs

\ni 
We also rewrite the identity operator in the auxiliary field propagators

\be
{ \bar{\cal D}^2 {\cal D}^2 + {\cal D}^2 \bar{\cal D}^2 - 
 {\cal D} \bar{\cal D}^2 {\cal D} - m \bar{m} 
 \over \Box - m \bar{m} } = 1
\ee 

\ni 
and we find

\beqs
 \lefteqn{ \langle \tU (1) \; \btU (2) \rangle \mid_{\th_{\a}^2=0} \; =  
 \; {\zZ^{2p-1} - \zl^{2p-1} \over \zZ^{2p} } 
 \( {m {\cal D}^2 \over \Box - m \bar{m} } - 
 \zl \zZ {\bar{m} \bar{\cal D}^2 \over \Box - m \bar{m} } \)  
 \d^4 (\th_{12}) \d^4 (x_{12}) \label{complex_propp}}\\ \nn \\
& & \qquad - {\zl^{2p-1} - \zZ^{2p-1} \over \zZ^{2p} (\zl-\zZ)} 
  \( \zZ^2 { {\cal D}^2 \bar{\cal D}^2 \over \Box - m \bar{m}} + 
    \zl^2 {\bar{\cal D}^2 {\cal D}^2 \over \Box - m \bar{m} } - 
    \zl \zZ ( {\cal D} \bar{\cal D}^2 {\cal D} + m \bar{m}) \) 
  \d^4 (\th_{12}) \d^4 (x_{12}) \nn \ .
\eeqs 

\ni 
Similarly the (ant)arctic hypermultiplet has the following reduced
form in $N=1$ superspace 

\beqs
 \langle \tU (1) \; \btU (2) \rangle  \mid_{\th_{\a}^2=0} & = &
 \langle \tilde{\U}_0 (1) \bar{\tilde{\U}}_0 (2) \rangle \ + 
 \zl \ \langle \tilde{\U}_1 (1) \bar{\tilde{\U}}_0 (2) \rangle -
 {1 \over \zZ} \langle \tilde{\U}_0 (1) \bar{\tilde{\U}}_1 (2) \rangle
 + \nonumber \\
& & + \(\zZl\) \langle \tilde{\U}_1 (1) \bar{\tilde{\U}}_1 (2) \rangle + 
 \sum_{n=2}^{+\infty} \(-\zZl\)^n \langle \tilde{\U}_n (1) 
 \bar{\tilde{\U}}_n (2) \rangle            \nn 	  \\ 
& = & \left[ - { {\cal D}^2 \bar{\cal D}^2 \over \Box - m \bar{m}} - 
 \zl {\bar{m} \bar{\cal D}^2 \over \Box - m \bar{m} } + 
 {1 \over \zZ} {m {\cal D}^2 \over \Box - m \bar{m} } \right. \nn \\ 
& & \left. - \zZl \(1 -{\bar{\cal D}^2 {\cal D}^2 \over \Box - m \bar{m}} \) 
  - \(\zZl\)^2 \sum_{n=0}^{+\infty} \(\zlZ\)^n \right]
 \d^4 (\th_{12}) \d^4 (x_{12}) \nn \\  \nn \\
& = & \left[ - {1 \over \zZ^2} \sum_{n=0}^{+\infty} \(\zZl\)^n 
 { \zl^2 \bar{\cal D}^2 {\cal D}^2 + \zZ^2 {\cal D}^2 \bar{\cal D}^2 
 - \zl \zZ {\cal D} \bar{\cal D}^2 {\cal D} 
 - \zl \zZ m \bar{m} \over \Box - m \bar{m} } \right.  \nn \\ \nn \\
& & \left. - \zl {\bar{m} \bar{\cal D}^2 \over \Box - m \bar{m} } + 
 {1 \over \zZ} {m {\cal D}^2 \over \Box - m \bar{m} } \right]
 \d^4 (\th_{12}) \d^4 (x_{12}) \ .
\label{Ups_prop}
\eeqs

\ni
When $|\zl /\zZ| < 1$ this result is consistent with the limit 
$2p \rightarrow +\infty$ of (\ref{complex_propp}). Not surprisingly
we find the same convergence problems as in the massless case
\cite{hyper_feynman}. 
  
We also expect the $N=2$ propagator to be 
proportional to $\na^4_1 \na^4_2 \d^8_{12}$, but now we have to be 
careful with the central charges of the algebra. We assign a global 
$U(1)$ charge $q=1$ to the {\em covariantized} spinor derivatives 
(to be justified later 
when we derive the massive propagator in $N=2$ superspace) 

\beqs
\tna^2 = {\cal D}^2 + \z {{\cal D} {\cal Q} + {\cal Q} {\cal D} \over 2} 
 + \z^2 {\cal Q}^2 = {\cal D}^2 + \z ({\cal D}{\cal Q} + \bar{m}) + 
 \z^2 {\cal Q}^2 = {\cal D}^2 + \z ({\cal Q}{\cal D} - \bar{m}) 
 + \z^2 {\cal Q}^2 & & \nn \\
& & \\
\btna^2 = \bar{\cal Q}^2 - \z {\bar{\cal D} \bar{\cal Q} + 
 \bar{\cal Q}\bar{\cal D} \over 2} + \z^2 \bar{\cal D}^2 = 
 \bar{\cal Q}^2 - \z (\bar{\cal D} \bar{\cal Q} - m) 
  + \z^2 \bar{\cal D}^2 = \bar{\cal Q}^2 - \z (\bar{\cal Q} \bar{\cal D} + m) 
  + \z^2 \bar{\cal D}^2 \nn  & . & 
\eeqs

\ni
Let us consider the $N=1$ projection of the following expression

\beqs
\tna^4_1 \tna^4_2 \d^8 (\th_{12}) \mid_{\th_{\a}^2=0} & = & 
 (\zZ-\zl)^2 \( \zl^2 \bar{\cal D}^2 {\cal D}^2 + 
  \zZ^2 {\cal D}^2 \bar{\cal D}^2 
 - \zl \zZ {\cal D} \bar{\cal D}^2 {\cal D} - \zl \zZ m \bar{m} \) 
 \d^4 (\th_{12}) \nn \\  
& & - m (\zZ-\zl)^3 {\cal D}^2 \d^4 (\th_{12}) + 
 \bar{m} \zl \zZ (\zZ-\zl)^3 \bar{\cal D}^2 \d^4 (\th_{12})  \ .
\label{n1_hyper_proj}
\eeqs

\ni
This suggests that the correct massive $N=2$ propagator for the complex 
$O(2p)$ hypermultiplet is the naive generalization of the massless
propagator 

\be 
 \langle \tU (1) \; \btU (2) \rangle = 
 - {\zl^{2p-1} - \zZ^{2p-1} \over \zZ^{2p} (\zl-\zZ)^3 } 
 {\tna^4_1 \tna^4_2 \over \Box - m \bar{m} } \d^8 (\th_{12})
 \d^4 (x_{12}) \ ,
\label{obvious}
\ee

\ni
whereas for the (ant)arctic hypermultiplet it is

\be 
 \langle \tU (1) \; \btU (2) \rangle = 
 - {1 \over \zZ^2}  \sum_{n=0}^{+\infty} \(\zZl\)^n 
 {\tna^4_1 \tna^4_2 \over (\zl -\zZ)^2 (\Box - m \bar{m}) } \d^8 (\th_{12})
 \d^4 (x_{12}) \ .
\label{arctic_prop}
\ee

\ni 
Note that if we take the convergent limit of the infinite sum also the
last two terms in (\ref{Ups_prop}) are reproduced by the $N=1$
projection of this expression. As in the massless case, we have to 
conjecture that this 
limit can be analytically continued to the region of no convergence.

The conjugate propagators are obtained from the projection of

\be
 \langle \btU (1) \; \tU (2) \rangle =  
 - {\zZ^{2p-1} - \zl^{2p-1} \over \zl^{2p} (\zZ-\zl)^3 } 
 {\tna^4_1 \tna^4_2 \over \Box - m \bar{m} } \d^8 (\th_{12})
 \d^4 (x_{12}) \ , 
\ee

\ni
and

\be
 \langle \btU (1) \; \tU (2) \rangle = 
 - {1 \over \zl^2}  \sum_{n=0}^{+\infty} \(\zlZ\)^n 
 {\tna^4_1 \tna^4_2 \over (\zl -\zZ)^2 (\Box - m \bar{m}) } \d^8 (\th_{12})
 \d^4 (x_{12}) \ .
\ee

\ni
where, to get the correct $N=1$ projection, we must use supercovariant 
derivatives obeying the algebra (\ref{n2_algebra}) with $q=-1$.

\subsection{Calculation of the massive hypermultiplet propagator in 
$N=2$ superspace}

To derive the massive hypermultiplet propagator in $N=2$ superspace we
must couple this multiplet to (ant)arctic sources. We consider the 
charged ordinary hypermultiplet interacting with an Abelian real
tropical multiplet which will develop a {\em v.e.v.}. The $N=2$ and
gauge invariant free action with sources that we want is the following

\be
 S_0+S_J = \int dx \; d^4 \th \oint {d \z \over 2 \pi i \z }  \bU
 e^V \U + \int dx \; d^8 \th \oint {d \z \over 2 \pi i \z } 
 \( \bar{J} e^V \U +  \bar{\U} e^V J \) \ .
\ee
 
\ni
The source $J$ is an unconstrained arctic field transforming as 

\be
J' = e^{i \Lambda } J 
\ee

\ni
with charge $q=1$ under the $U(1)$ symmetry (\ref{gauge_sym}). The field 
$e^V J$ can be regarded as a complex tropical source transforming with
the same charge but with the antarctic gauge parameter
$\bar{\Lambda}$. Writing the action with {\em redefined} (ant)arctic 
sources $J \longrightarrow e^{V_+} J$ and the covariantly
projective (ant)arctic hypermultiplets (\ref{covariant_arctic}), we have

\be
 S_0+S_J = \int dx \; d^4 \th \oint {d \z \over 2 \pi i \z }  \btU \tU
 + \int dx \; d^8 \th \oint 
 {d \z \over 2 \pi i \z } \( \bar{J} \tU + \btU J \) \ .
\ee

\ni
If we use the covariantly projective complex $O(2p)$ hypermultiplet
(\ref{covariant_2p}) we obtain an action of exactly the same form. The
difference is that the $\tU$ fields are of finite order in $\z$ and
thus the contour integration forces all but a finite number of
component sources to decouple. Reducing the corresponding two actions 
to $N=1$ components we recover (\ref{n1_source_act}) and its equivalent 
for the $O(2p)$ multiplet.

We want to write the kinetic action and source terms as integrals
with the full $N=2$ superspace measure, and integrate out the 
hypermultiplet in the free theory path integral. We analyze the 
complex $O(2p)$ hypermultiplet since the (ant)arctic one is simply
reproduced by taking the limit $2p \rightarrow +\infty$. As in the
massless case \cite{hyper_feynman} we use an unconstrained 
$O(2p-4)$ prepotential 

\be
 \tilde{\U} = e^{V_+} \na^4 e^{-V_+} \( e^{V_+}\psi \)
  = \tna^4 \sum_{n=0}^{2p-4} \tilde{\psi}_n \z^n, \;\; 
 \bar{\tilde{\U}} = e^{V_-} {\na^4 \over \z^4} e^{-V_-} 
 \( e^{V_-} \bar{\psi} \) = {\tna^4 \over \z^4}
 \sum_{m=0}^{2p-4} { \bar{\tilde{\psi}}_{m} \over (-\z)^m} \ .
\ee

\ni
The action with sources written in terms of $\tilde{\psi}$
is\footnote{ Note that after doing the integration by parts in the
second source term all covariantized projective derivatives have 
the same $U(1)$ charge $q=1$.}

\be
 S_0+S_J = \int dx \; d^8 \th \oint {d \z \over 2 \pi i \z } 
 \(\bar{\tilde{\psi}} { e^{V_+} \na^4 e^{-V_+} \over \z^2} \tilde{\psi} 
 + \bar{J} e^{V_+} \na^4 e^{-V_+} \tilde{\psi} + 
 \bar{\tilde{\psi}} {e^{-V_-} \na^4 e^{V_-} \over \z^4}  J \) \ .
\label{analy-source}
\ee

To integrate out 
the prepotential we use the techniques introduced in the massless
case: we insert a projector in the source term so that we can factor
out the kinetic operator and complete squares by shifting the
prepotential with a $O(2p-4)$ polynomial. From the algebra
(\ref{na_D_algebra1}-\ref{na_D_algebra3}) we learn that the projector 
operator for $q=1$ fields living in projective superspace is given by

\be
{\tna^4 \tD^4 \over 16 (\Box - m \bar{m})^2 } \tna^4 = \tna^4 \ .
\ee

\ni
We are also free to add to $\tD^4$ any operator annihilated by the
$\tna^4$ to the left and to the right. It is particularly convenient to
choose the following combinations

\beqs
 K = { \( \tD^2 - {1 \over \z} \tna \tD + {2 \over \z} \bar{m} + 
 {1 \over \z^2} \tna^2 \) \( \btD^2 -{1 \over \z} \btD \btna 
 - {2 \over \z} m + {1 \over \z^2} \btna^2 \) \over 16(\Box-m\bar{m})^2 } 
 = { {\cal Q}^2 \bar{\cal D}^2 \over (\Box-m\bar{m})^2} \ , \nn  \\
 L(\z) = { \( \tD^2 + {1 \over \z} \tna \tD - {2 \over \z} \bar{m} + 
 {1 \over \z^2} \tna^2 \) \( \btD^2 + {1 \over \z} \btD \btna 
 + {2 \over \z} m + {1 \over \z^2} \btna^2 \) \over 16(\Box-m\bar{m})^2} = 
 { {\cal D}^2 \bar{\cal Q}^2 \over \z^4 (\Box-m\bar{m})^2} \ ,
\eeqs

\ni
which satisfy the following useful relation

\be
 \tna^{4}(\z) L(\z') \tna^{4}(\z') = \( {\z \over \z'} \)^2
   \tna^{4}(\z) K \tna^4(\z')  
\label{nice_relation}
\ee

\ni
Using this operators, the last source term can be rewritten

\be
  \oint {d \z \over 2 \pi i \z } \bar{\tilde{\psi}} {\tna^4 \over \z^2}
  K  {\tna^4 \over \z^2}J \ .
\label{common_steps}
\ee

\ni
The differential operator $\bar{\tilde{\psi}} \tna^4 / \z^2$ is a
$O(-2p+2,2)$ polynomial in $\z$. However the operator $K$ 
annihilates the coefficients of $\z$ and $\z^2$ in $\tna^4 /\z^2$, and
therefore only the $O(-2p+2,0)$ piece contributes to the source
coupling. This means that we can project the expression multiplying 
$\bar{\tilde{\psi}} \tna^4 / \z^2$ onto the subspace of $O(0,2p-2)$ 
polynomials. Using the delta functions on the Riemann sphere that we
defined in \cite{hyper_feynman}, we rewrite the source term

\be
 \oint {d \z \over 2 \pi i \z } \bar{\tilde{\psi}} {\tna^4 \over \z^2}
  \oint {d \z' \over 2 \pi i \z' } \d_{(0)}^{(2p-2)} (\z, \z')  
 K {\tna^4 (\z') \over \z'^2} 
 J(\z') = \oint {d \z \over 2 \pi i \z } \bar{\tilde{\psi}} 
 {\tna^4 \over \z^2} {\cal J} (\z) \ ,
\ee

\ni
where we have defined the source

\be
{\cal J}(\z) = \oint {d \z' \over 2 \pi i \z' } 
 \left( \d_{(0)}^{(2p-4)}(\z,\z')K +
     \d_{(2p-5)}^{(2p-4)}(\z,\z')L(\z') \right)
  {\tna^{4}(\z') \over \z^{\prime 2}} J(\z')
\ee

\ni
which is a $O(0,2p-4)$ superfield, and due to the identity
(\ref{nice_relation}) satisfies 

\be
 \tna^{4}{\cal J} = \oint {d \z' \over 2 \pi i \z' }
  \d_{(0)}^{(2p-2)}(\z,\z')K
   {\tna^{4}(\z') \over \z^{\prime 2}} J(\z') \ .
 \label{J_rel}
\ee

\ni
Similarly we may rewrite the conjugate source term as

\be
 \oint {d \z \over 2 \pi i \z}
  \tilde{\psi}\tna^4\bar{J} =
 \oint {d \z \over 2 \pi i \z}
   \tilde{\psi}{ \tna^4 \over \z^2 } \bar{\cal J} \ ,
\ee

\ni
where we have used the fact that the operator $L$ annihilates the
coefficients of the $\z^{-2}$ and $\z^{-1}$ terms in the operator
$\tna^{4}/\z^2$, and where we have defined

\be
 \bar{\cal J} = \oint {d \z' \over 2 \pi i \z' }
  \left( \d_{(4-2p)}^{(0)}(\z,\z') L(\z') + 
   \d_{(4-2p)}^{(5-2p)}(\z,\z') K\right)
  \z^{\prime 2} \tna^{4}(\z') \bar{J}(\z')
\ee

\ni
satisfying

\be
 \tna^{4}\bar{\cal J} = \oint {d \z' \over 2 \pi i \z' }
  \d_{(2-2p)}^{(0)}(\z,\z') L(\z') \z^{\prime 2} \tna^{4} (\z') 
 \bar{J}(\z') \ .
\label{bar_J_rel}
\ee

\ni
The new sources ${\cal J},\bar{\cal J}$ have the correct order in
$\z$, so we may complete squares in the action
  
\be
 S_0 + S_J = \int dx \; d^8 \th \oint {d \z \over 2 \pi i \z } 
 \( \bar{\tilde{\psi}} + \bar{\cal J} \) { \tna^4 \over \z^2} 
 \( \tilde{\psi} + {\cal J} \) - \bar{\cal J} { \tna^4 \over \z^2} 
 {\cal J} \ .
\ee

\ni
Integrating out the hypermultiplet prepotential in the free theory 
path integral, we are left with the term quadratic in sources. 
Following our previous arguments backwards and using the equations
(\ref{nice_relation},\ref{J_rel},\ref{bar_J_rel}), we find

\beqs
 ln Z_0 [J, \bar{J}] & = & - \int dx \; d^8 \th \oint {d \z \over 2 \pi i \z }
  \bar{\cal J} { \tna^4 \over \z^2} {\cal J}   \label{path_integr} \\
& = & - \int dx \; d^8 \th \oint {d \z \over 2 \pi i \z}  \bar{J} \tna^4  
 {\cal J} \nn \\ 
& = & - \int dx \; d^8 \th \oint {d \z \over 2 \pi i \z}  
 \oint {d \z' \over 2 \pi i \z' } \bar{J} (\z) \tna^4 (\z)
 \d_{(0)}^{(2p-2)} (\z, \z')  
 K {\tna^4(\z') \over \z'^2} J(\z') \ . \nn 
\eeqs

The $N=2$ propagator is obtained by functional differentiation
with respect to the unconstrained sources that couple to the 
hypermultiplet

\beqs
 \langle \tU (1) \btU (2) \rangle & = & 
 {\d \over \d \bar{J} (x_1, \th_1,\z_1)} {\d \over \d J
 (x_2,\th_2,\z_2)} \ln Z_0           \nonumber \\ \nonumber \\
& = &  -\oint {d\z \over 2 \pi i \z} \oint {d\z' \over 2 \pi i \z'} \;
  \d_{(0)}^{(2p)}(\z_1,\z) \; \d_{(-2p)}^{(0)}(\z_2,\z') \;
  \d_{(0)}^{(2p-2)}(\z,\z')                         \\
& & \qquad\qquad\qquad \times\quad  
 {\tna^{4} (\z) K \tna^{4} (\z') \over \z'^{2} (\Box - m \bar{m})^2 } \;
  \d^{8}(\th_1 - \th_2) \; \d(x_1-x_2) \ . \nn
\eeqs

\ni
Since the operator $K$ annihilates the two highest order coefficients
in $\tna^4 (\z)$ and $\tna^4 (\z')$, the product 

\be
 \d_{(0)}^{(2p-2)}(\z,\z') \tna^4 (\z) K { \tna^{4} (\z') \over \z'^2}
\ee

\ni 
is $O(0, 2p)$ in $\z$ and $O(-2p,0)$ in $\z'$. Hence, the integration 
of the complex coordinates gives the following propagator

\be
 \langle \tU (1) \btU (2) \rangle = - \d_{(0)}^{(2p-2)}(\zl,\zZ) 
 {\tna_1^4 K \tna_2^4  \over \zZ^2 (\Box - m \bar{m})^2 } \;
  \d^{8}(\th_1 - \th_2) \; \d(x_1-x_2) \ .
\label{positive_q}
\ee

\ni
To obtain the form of the propagator proposed above, we must use the
following identities

\be
 \tna_1^2 {\cal Q}^2 = {\cal D}^2 {\cal Q}^2 + \zl \bar{m} {\cal Q}^2 = 
{ {\tna_1^2  \tna_2^2 \over \zZ-\zl} - \bar{m} \tna_1^2 \over \zZ-\zl} \ ,
\ee

\be
 \bar{\cal D}^2 \btna_2^2 = \bar{\cal D}^2 \bar{\cal Q}^2 + 
 \zZ m \bar{\cal D}^2 = { {\btna_1^2 \btna_2^2 \over \zZ-\zl} + 
 m \btna_2^2 \over \zZ-\zl} \ .
\ee

\ni
Using the anticommutator $\{ \tna_1^\a , \btna_2^{\db} \} = i (\z_1
- \z_2) \pa^{\a \db}$ they allow us to rewrite the numerator in 
(\ref{positive_q})

\be
 \btna_1^2  
\({\tna_1^2 \tna_2^2 \over (\zZ-\zl)^2} - { \bar{m} \tna_1^2 \over \zZ-\zl}\)
\({\btna_1^2 \btna_2^2 \over (\zZ-\zl)^2} + { m \btna_2^2  
 \over \zZ-\zl} \) \tna_2^2 = 
 {\Box - m \bar{m} \over (\zZ-\zl)^2 } \tna_1^4 \tna_2^4 \ .
\ee

\ni
The propagator may now be written as follows

\be
 \langle \tU (1) \btU (2) \rangle = - \d_{(0)}^{(2p-2)}(\zl,\zZ) 
 {\tna_1^4 \tna_2^4  \over \zZ^2 (\zZ-\zl)^2 (\Box - m \bar{m}) } \;
  \d^{8}(\th_1 - \th_2) \; \d(x_1-x_2) \ ,
\ee

\ni
and using the explicit from of the Riemann sphere delta function 
we recover the expression (\ref{obvious}) proposed above. 

To obtain the  conjugate propagator we integrate by parts in
(\ref{path_integr}), which has the effect of reversing the sign in
the $U(1)$ charge of the covariantized derivative\footnote{ This is
the natural sign for the derivatives acting on $\bar{J}$ which has the
opposite $U(1)$ charge.}, and we take the corresponding functional
derivatives with respect to unconstrained sources. This explains the
previous $U(1)$ charge assignment in the covariant derivatives of the
conjugate propagator. Using the same integration by parts we deduce that 
the $U(1)$ charge changes when we use transfer rules

\be
\tna_2^4 (q=1) \d^8 (\th_1 - \th_2) = \d^8 (\th_1 - \th_2)
 \stackrel{\longleftarrow}{\tna_2^4} (q=-1) \ .
\ee

The (ant)arctic multiplet propagator is obtained as the
$2p \rightarrow \infty$ limit of the above expressions. The
arguments are somewhat simplified as in the massless case because of
the infinite order in the coordinate $\z$. The final result reproduces
our ansatz (\ref{arctic_prop}).

\section{Feynman rules for diagram construction}

The rules for constructing diagrams are very similar to those used
with the massless hypermultiplet \cite{hyper_feynman}. For
completeness we briefly summarize them here:

\begin{itemize}
\item
Propagators:\\
 We put the $\tna^{4}$ factors of the propagators in the vertices. With
this convention the propagator of the complex $O(2p)$ multiplet becomes
 
\be
  \langle \tU(1)\bar{\tU}(2)\rangle =
  -\frac{\d_{(0)}^{(2p-2)}(\z_1,\z_2)}{\z_{2}^{2}
    (\z_1-\z_2)^{2}(\Box-m\bar{m})} \ ,
 \ee

\ni
and its conjugate
 
\be
  \langle \bar{\tU}(1)\tU(2)\rangle =
  -\frac{\d_{(2-2p)}^{(0)}(\z_1,\z_2)}{\z_{1}^{2}
    (\z_1-\z_2)^{2}(\Box-m\bar{m})} \ .
 \ee

\ni
The (ant)arctic multiplet propagators are obtained from the
propagators above by letting $p\longrightarrow\infty$.

\item
Vertices:\\ These are read directly from the interaction Lagrangian. It
is most convenient to put the spinor derivatives $\tna^{4}_{1},
\tna^{4}_{2}$ of a propagator in the vertices. This means that every
internal line of a vertex will have a $\tna^{4}$ factor except one
where the spinor derivative has been used to complete the restricted
superspace measure of the interaction vertex to a full $N=2$ 
superspace measure.

\item
We may always absorb in the Grassmann measure of a projective vertex
a factor of $\tna^{4}$ from an internal line, so we will integrate 
over $d^{8}\th$ at each vertex. There is
also an integral over the $x$-space position of each vertex, or,
equivalently, an integral over each loop momentum plus an overall
factor $\propto \d(\S k_{ext})$.

\item In computing any particular diagram, we amputate the external 
line propagators, which means that there are no $\tna^4$ factors on
external lines.

\item Finally, there may be symmetry factors associated with certain
graphs and they are calculated in the usual fashion. The
``$D$''-algebra is performed and the Grassmann coordinate dependence
of the propagators is reduced until we obtain a local expression in
Grassmann space.

\end{itemize}

\section{Acknowledgements}
The authors would like to thank Martin Ro\v{c}ek for helpful
discussions. RvU would like to thank FGR for his hospitality while
completing this work.

\end{document}